# The uncharted space of non-Hermitian solutions to the Hartree-Fock and Kohn-Sham equations

Matthias Ernzerhof[*] and Mohamed Loutis

*Département de Chimie, Université de Montréal,*

*C.P. 6128 Succursale A, Montréal, Québec H3C 3J7, Canada*

Pierre-Olivier Roy

*Department of Physical Chemistry, University of Geneva, CH-1211 4 Genève, Switzerland*

Didier Mayou

*Institut Néel, 25 Avenue des Martyrs,*

*BP 166, 38042 Grenoble Cedex 9, France*

(Dated: October 29, 2025)

## Abstract

Many problems in physical chemistry involve systems that are coupled to an environment, such as a molecule interacting with an adjacent surface, possibly resulting in meta-stable molecular states where electron density is transferred to the surface. Such systems can be described by non-Hermitian quantum mechanics (NHQM), where the Hamiltonian includes a complex potential. Within NHQM, one can also formulate the Hartree-Fock (HF) and Kohn-Sham (KS) methods and, as in the conventional theory, an effective independent-particle picture is employed. The crucial observation of the present work is that even for systems that are not coupled to an environment, in the non-Hermitian HF or KS equation a single electron is interacting with the remaining electrons, which act as an environment, allowing for the exchange of current density between the one-electron and the remaining $(N-1)$-electron system. The corresponding self-consistent states represent a new uncharted space of solutions to the HF and KS equations. We show that the additional solutions can have a physical interpretation and thus extend the range of problems HF and KS can be applied to. If open-system HF and KS calculations are performed, the new class of solutions is always encountered but this has also not been noted previously.

---

[*] matthias.ernzerhof@gmail.com

## I. INTRODUCTION

Non-Hermitian quantum mechanics (NHQM) [1] is an extension of conventional quantum mechanics within which the Hermiticity of the Hamiltonian is not imposed. Non-Hermitian (NH) Hamiltonians ($H \neq H^\dagger$) naturally arise in the description of open quantum systems [2], where probability density can be exchanged with the environment. To provide a simple explicit example, we focus on the source-sink potential approach (SSP) [3–11], summarized in Fig. 1. SSP is an elementary model that we previously developed for molecules coupled to external contacts. In the simplest version of SSP, the Hückel Hamiltonian of a diatomic

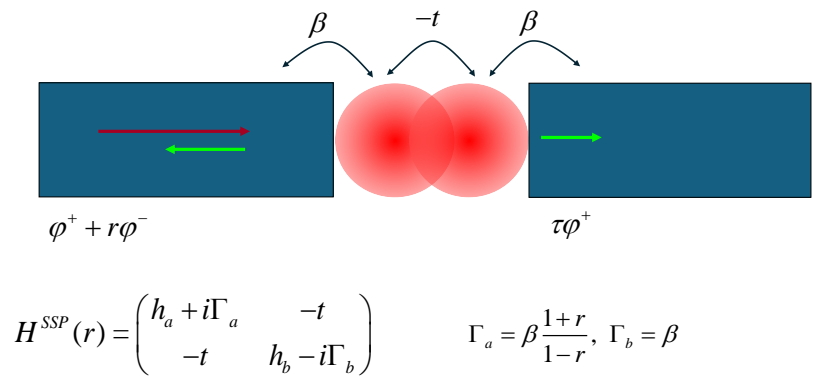

FIG. 1. Summary of the source-sink potential (SSP) model for molecular conductance. In the left contact, represented by a rectangle, an incoming electron wave $\varphi^+$ of energy $\epsilon$ is partially reflected into a back-traveling one $\varphi^-$ of the same energy with reflection coefficient $r(\epsilon)$. The electron transmitted through the diatomic molecule continues in the right contact in a forward going wave with an amplitude $\tau(\epsilon)$. The transmission probability $T(\epsilon)$ is related to the conductance $g(\epsilon)$ through $g = g_0 T(\epsilon), T(\epsilon) = |\tau(\epsilon)|^2 = 1 - |r(\epsilon)|^2$. $g_0 = \frac{e^2}{h}$ is the conductance quantum, with electron charge $e$ and Planck constant $h$. This infinite system can be replaced by a finite one with the help of a non-Hermitian Hamiltonian. In the Hückel version of SSP, the external contacts are accounted for through complex potentials $h_a + i\Gamma_a$, $\Gamma_a = \beta\frac{1+r}{1-r}$ and $h_b - i\Gamma_b$, $\Gamma_b = \beta$, respectively, which appear in the molecular Hamiltonian matrix $H^{SSP}(r(\epsilon))$. The coupling parameter between the molecule and contacts is given by $\beta$ and the molecular coupling parameter is $-t$. $r(\epsilon)$ is then obtained through the condition that the real energy of the incoming electron $\epsilon$ is reproduced by $H^{SSP}(r(\epsilon))$, i.e., $det(H^{SSP}(r(\epsilon)) - \epsilon) = 0$.

molecule coupled to two external contacts is,

$$H^{\text{SSP}} = \begin{pmatrix} h_a + i\Gamma_a & -t \\ -t & h_b - i\Gamma_b \end{pmatrix}, \quad (1)$$

$-t$ is the real-valued hopping term between the two atoms $a$ and $b$, each being coupled to a semi-infinite contacts through which electron current density can be inserted and extracted, respectively. The complex diagonal elements $(h_a + i\Gamma_a, h_b - i\Gamma_b)$ are determined such that they describe the current density injection/extraction of a particular physical process of interest. These processes include stationary electron transport through the molecule. This case is of particular importance here and we preempt a key result obtained below. With a suitable choice of parameters we obtain

$$H^{\text{SSP}} = \begin{pmatrix} i & -1 \\ -1 & -i \end{pmatrix}, \quad \text{where} \quad h_a = h_b = 0, \Gamma_a = \Gamma_b = 1, t = 1. \quad (2)$$

Referring to Fig. 1, $\Gamma_a = \Gamma_b = 1$ corresponds to $r = 0$ and $\beta = 1$. The matrix Eq. 2 describes electron-density current entering a diatomic molecule on atom $a$ and leaving on atom $b$ after passing through the molecule. Below, we study self-consistent solutions for an isolated, two-electron, diatomic molecule and find that $H^{\text{SSP}}$ in Eq. 2 is essentially identical to a Fock matrix of this system. This is because the mean field created by the second electron in the Fock matrix mimics the environment provided by the external contacts in the SSP matrix.

In previous work (see for example, [7, 9, 12–18]), ground-state density functional theory (DFT) and also Hartree-Fock (HF) theory have been extended and applied to the realm of NHQM. In these methods, electron interaction effects are considered for problems of molecules coupled to an environment. The results reported depend of course on the nature of the mean-filed approximation employed. Here we investigate the case where even for an isolated molecule, NH Fock and Kohn-Sham (KS) matrices appear as solutions of the self-consistent equations. Furthermore, if KS or HF calculations are performed within the NH formalism to describe open systems, the new class of solutions described in this work is always present, but it has not been previously noted.

Recently, NHQM has been receiving considerable attention. In particular, NH parity-time (PT) symmetric systems [19] are investigated, where the Hamiltonian is invariant under combined parity and time-reversal operations. Such Hamiltonians can have real spectra and are considered for the description of physical phenomena. Furthermore, there are numerous applications of NHQM in electronic structure theory, examples include complex absorbing potentials [12, 18, 20, 21], complex scaling [1, 22], and NH self-energies in Green's-function (electron-propagator) methods [23].

Closely related to PT symmetric NHQM and to the present work are the articles by Burton, Thom, Loos and collaborators [24–28] which focus on NH HF and KS theory and extend them through analytic continuation of the orbital coefficients into the complex plane. Here we also advocate to consider the entire class of additional solutions that are provided by NH theory, including those with complex energies. We provide a physical interpretation of these solutions, the energies, orbitals, and the corresponding mean-field potentials, by relating them to open quantum systems. Furthermore, we show that NH HF and KS theories of isolated systems are a special case of our previously developed open-system HF and KS theories. Next we summarize some relevant aspects of NHQM, before considering NH mean-field theory.

## II. NON-HERMITIAN QUANTUM MECHANICS

Given a diagonalizable, NH Hamiltonian $H$, its right eigenvectors $|\psi_n\rangle$ and left eigenvectors $\langle\phi_n|$ satisfy

$$H|\psi_n\rangle = E_n|\psi_n\rangle, \ \langle\phi_n|H = E_n\langle\phi_n|. \tag{3}$$

Non-Hermiticity introduces modifications of the fundamental equations of quantum mechanics, such as the variational principle, where a left and a right trial function are introduced in the energy expression

$$E = \frac{\langle\phi|H|\psi\rangle}{\langle\phi|\psi\rangle}. \tag{4}$$

For complex symmetric Hamiltonians, which are of interest here, the left eigenvector is the complex conjugate of the right eigenvector, $\phi = \psi^*$. Furthermore, we introduce the $C$-scalar product $(\phi|\psi) = \langle\phi^*|\psi\rangle$, leading to the variational principle

$$\begin{aligned}\delta E &= \delta \frac{(\psi|H|\psi)}{(\psi|\psi)} \\ &= 0.\end{aligned} \tag{5}$$

The solutions to the stationary principle in Eq. 5 are those of the Schrödinger equation. It is crucial to note that the NH version of the expectation value of an operator $O$, as implicit in Eq. 5, i.e.,

$$(O) = \frac{(\psi|O|\psi)}{(\psi|\psi)}, \quad \text{non-Hermitian expectation value}, \tag{6}$$

results in an electron density $\rho(\mathbf{r}) = \frac{(\psi|\hat{\rho}|\psi)}{(\psi|\psi)}$ that is complex in general. For instance, in NHQM the electron density of an orbital is $\rho(\mathbf{r}) = \psi(\mathbf{r})\psi(\mathbf{r})/(\psi|\psi)$. Similarly, for a many-electron system, the electron density is obtained through

$$\rho^\sigma(\mathbf{r}) = N \int d2 \ldots dN \, \frac{\psi(\mathbf{r}, \sigma, 2, \ldots, N)\, \psi(\mathbf{r}, \sigma, 2, \ldots, N)}{(\psi|\psi)}, \tag{7}$$

where $\sigma$ is a spin coordinate and $2, 3, \ldots, N$ are combined spin and spatial coordinates. The stationarity principle in Eq. 5 provides the basis for the generalization of HF and KS (DFT) to the NH domain [7, 13, 16]. The Hohenberg-Kohn theorems have been adapted [13] to accommodate complex densities, ensuring a limited one-to-one correspondence between external potentials and electron densities. Additionally, the Levy constrained search method has been expanded [7] to handle complex densities. In a nutshell, in complex-density functional theory (CODFT), the conventional expectation value is replaced by its NH version and the formulas of DFT, including KS theory, remain valid within specified limitations. Similarly, in HF theory, the one-particle density matrix is replaced by one calculated using the NH left and right eigenfunctions.

To further illustrate the physical implications of complex symmetric Hamiltonians, we consider an example of the form

$$H = H_0 - iw, \tag{8}$$

where $H_0$ is a real, Hermitian operator and $w(\mathbf{r})$ is a real potential. The resulting eigenvalues are generally complex, $E = E_0 - i\frac{\Gamma}{2}$, where $E_0$ and $\Gamma$ are real constants. A stationary state with such a complex eigenvalue has the time dependence

$$\psi(t) = \psi\, e^{-iE_0 t} e^{-(\Gamma/2)t}, \tag{9}$$

which shows exponential decay for positive $\Gamma$.

The eigenstates of the symmetric, NH Hamiltonian provide the natural basis for the description of a given open system. The corresponding complex eigenvalues determine resonance positions and exponential decay (or growth) rates, but do not represent the real valued measurable energy corresponding to $\psi$. Instead, the physical energy $E^{\text{H}}$ of a given wave function $\psi$ is obtained using the conventional expectation value of $H_0$, the underlying Hermitian Hamiltonian,

$$E^{\mathrm{H}} = \frac{\langle \psi | H_0 | \psi \rangle}{\langle \psi | \psi \rangle}. \tag{10}$$

We stress that in general, $\operatorname{Re} E \neq E^H$.

For a Hamiltonian of the form $H = H_0 - iw$, the continuity equation in the time-dependent domain is [13],

$$\frac{d\rho^H(\mathbf{r}, t)}{dt} = -\nabla \cdot \mathbf{j}^H(\mathbf{r}, t) - 2\, w(\mathbf{r}) \rho^H(\mathbf{r}, t) \tag{11}$$

which reduces for stationary states to

$$\Gamma \rho^H(\mathbf{r}) = \nabla \cdot \mathbf{j}^H(\mathbf{r}) + 2\, w(\mathbf{r}) \rho^H(\mathbf{r}). \tag{12}$$

In these equations the physical electron density $\rho^H$ and current densities $\mathbf{j}^H$ are calculated using the conventional expectation values. The important insight that Eqs. 11 and 12 convey is that the term $w(\mathbf{r})\rho(\mathbf{r})$ can add to or subtract from the divergence of the current density, i.e., to the amount of current density emerging (disappearing) from the point $\mathbf{r}$, and thus acts as a source (sink) for probability current density. Below we show that within the mean-field Hubbard approach, the Hartree term is a complex local potential if NH solutions to the mean-field equations are obtained and thus the Hartree-potential turns into a source or sink of current density, ensuring that all NH solutions have non-vanishing current densities.

## III. THE HUBBARD MODEL AND ITS HERMITIAN AND NON-HERMITIAN MEAN-FIELD APPROXIMATIONS

The Hubbard model is a useful tool in electronic structure theory for studying electron interaction. It captures the competition between kinetic energy (hopping of electrons between sites) and potential energy (on-site and electron-electron interaction terms). In general, the Hubbard Hamiltonian for a $n$-site system is given by,

$$H = \sum_{i=1,\sigma}^{n} h_i^\sigma \hat{n}_{i\sigma} - t \sum_{\langle i,j \rangle, \sigma} (\hat{c}_{i\sigma}^\dagger \hat{c}_{j\sigma} + \hat{c}_{j\sigma}^\dagger \hat{c}_{i\sigma}) + U \sum_{i=1}^{n} \hat{n}_{i\uparrow} \hat{n}_{i\downarrow}, \tag{13}$$

$h_i^\sigma$ is the on-site energy, $\hat{n}_{i\sigma} = \hat{c}_{i\sigma}^\dagger \hat{c}_{i\sigma}$ is the number operator. $\hat{c}_{i\sigma}^\dagger$ and $\hat{c}_{i\sigma}$ are the creation and annihilation operators for an electron with spin $\sigma$ at site $i$. $\langle i, j \rangle$ signifies the set of bonded sites, $t$ is the hopping parameter and $U$ is the on-site Coulomb interaction. In the illustrations below, we don't specify the units of the parameters, as only relative parameter values effect the results. In the MF approximation, the term $U \sum_{i=1}^{n} \hat{n}_{i\uparrow}\hat{n}_{i\downarrow}$ is decoupled by replacing a density operator with its expectation value,

$$H^{\text{HMF}} = \sum_{i=1,\sigma}^{n} h_i^\sigma \hat{n}_{i\sigma} - t \sum_{\langle i,j \rangle,\sigma} (\hat{c}_{i\sigma}^\dagger \hat{c}_{j\sigma} + \hat{c}_{j\sigma}^\dagger \hat{c}_{i\sigma})$$
$$+ U \left( \sum_{i=1}^{n} \langle \hat{n}_{i\uparrow} \rangle \hat{n}_{i\downarrow} + \hat{n}_{i\uparrow} \langle \hat{n}_{i\downarrow} \rangle - \langle \hat{n}_{i\uparrow} \rangle \langle \hat{n}_{i\downarrow} \rangle \right). \quad (14)$$

Here,

$$\langle \hat{n}_{i\sigma} \rangle = \sum_{l=1}^{occ} {}^l c_{i\sigma}^* \, {}^l c_{i\sigma} \quad (15)$$

is the conventional electron density on site $i$ which is calculated in terms of the occupied, orthonormalized spin orbitals (enumerated by $l$). In this Hermitian version of the MF operator, the conventional expectation value is used, ensuring that the electron density $\langle \hat{n}_{i\sigma} \rangle$ is real valued. In the Hubbard MF approach, there is no self-interaction contribution; therefore, if we interpret Eq. 14 as an approximation to the KS equation, only the Fermi and Coulomb correlation contribution to the total energy and KS potential are neglected.

To relax the constraint of Hermicity, we replace the conventional electron density $\langle \hat{n}_{i\sigma} \rangle$ by the density calculated using the NH expectation value (Eq. 6), which for orthonormalized $((\phi_i | \phi_j) = \delta_{ij})$ is

$$(\hat{n}_{i\sigma}) = \sum_{l=1}^{occ} {}^l c_{i\sigma} \, {}^l c_{i\sigma} \quad (16)$$

and obtain,

$$H^{\text{NHMF}} = \sum_{i=1,\sigma}^{n} h_i^\sigma \hat{n}_{i\sigma} - t \sum_{\langle i,j \rangle,\sigma} (\hat{c}_{i\sigma}^\dagger \hat{c}_{j\sigma} + \hat{c}_{j\sigma}^\dagger \hat{c}_{i\sigma})$$
$$+ U \left( \sum_{i=1}^{n} (\hat{n}_{i\uparrow})\hat{n}_{i\downarrow} + \hat{n}_{i\uparrow}(\hat{n}_{i\downarrow}) - (\hat{n}_{i\uparrow})(\hat{n}_{i\downarrow}) \right). \quad (17)$$

The crucial point illustrated below is that within NHQM, $(\hat{n}_{i\sigma})$ yields a complex-valued electron density for complex-valued orbitals. On the other hand a complex-valued $U(\hat{n}_{i\sigma})$ is the only possible source of non-Hermiticity in the NHMF model, it follows that NH solutions imply that the MF potential is complex and this in turn generates non-vanishing current densities.

We focus on a two site (diatomic) model and the orbitals are written as linear combinations of atomic basis functions ($\varphi_i$), with coefficients $a$ and $b$ for the up-spin and $c$ and $d$ for the down-spin orbital,

$$\phi_\uparrow = a\varphi_1 + b\varphi_2, \quad \phi_\downarrow = c\varphi_1 + d\varphi_2. \tag{18}$$

The energy expectation value for the non-Hermitian mean-filed (NHMF) method is given by,

$$E^{\text{NHMF}}(a,b,c,d) = \frac{(\Phi|H^{\text{NHMF}}|\Phi)}{(\Phi|\Phi)}, \tag{19}$$

where $\Phi$ denotes the Slater determinant composed of the up-and down-spin orbital. The explicit expression for the MF energy then becomes,

$$\begin{aligned} E^{\text{NHMF}}(a,b,c,d) &= \frac{1}{a^2+b^2} \begin{pmatrix} a & b \end{pmatrix} \begin{pmatrix} \ddots & & \\ & ^{\text{NH}}F^\uparrow_{ij} & \\ & & \ddots \end{pmatrix} \begin{pmatrix} a \\ b \end{pmatrix} \\ &+ \frac{1}{c^2+d^2} \begin{pmatrix} c & d \end{pmatrix} \begin{pmatrix} \ddots & & \\ & ^{\text{NH}}F^\downarrow_{ij} & \\ & & \ddots \end{pmatrix} \begin{pmatrix} c \\ d \end{pmatrix} \\ &- E^{\text{NH}}_{\text{Hartree}}, \end{aligned} \tag{20}$$

where

$$^{\text{NH}}F^\uparrow = \begin{pmatrix} h^\uparrow_a + \frac{c^2 U}{c^2+d^2} & -t \\ -t & h^\uparrow_b + \frac{d^2 U}{c^2+d^2} \end{pmatrix}, \quad ^{\text{NH}}F^\downarrow = \begin{pmatrix} h^\downarrow_a + \frac{a^2 U}{a^2+b^2} & -t \\ -t & h^\downarrow_b + \frac{b^2 U}{a^2+b^2} \end{pmatrix} \tag{21}$$

and

$$E^{\text{NH}}_{\text{Hartree}} = U \frac{(a^2 c^2 + b^2 d^2)}{(a^2+b^2)(c^2+d^2)}. \tag{22}$$

Here, $h^\sigma_a$ and $h^\sigma_b$, are the $\sigma$-spin on-site energies. This NHMF energy expression is contrasted to its Hermitian analog,

$$E^{\text{HMF}} = \frac{\langle \Phi | H^{\text{HMF}} | \Phi \rangle}{\langle \Phi | \Phi \rangle}$$

$$= \frac{1}{|a|^2 + |b|^2} \begin{pmatrix} a^* & b^* \end{pmatrix} \begin{pmatrix} \ddots & & \\ & {}^{\text{H}}F_{ij}^{\uparrow} & \\ & & \ddots \end{pmatrix} \begin{pmatrix} a \\ b \end{pmatrix}$$

$$+ \frac{1}{|c|^2 + |d|^2} \begin{pmatrix} c^* & d^* \end{pmatrix} \begin{pmatrix} \ddots & & \\ & {}^{\text{H}}F_{ij}^{\downarrow} & \\ & & \ddots \end{pmatrix} \begin{pmatrix} c \\ d \end{pmatrix}$$

$$- E_{\text{Hartree}}^{\text{H}}, \qquad (23)$$

where

$${}^{\text{H}}F^{\uparrow} = \begin{pmatrix} h_a^{\uparrow} + \frac{|c|^2 U}{|c|^2 + |d|^2} & -t \\ -t & h_b^{\uparrow} + \frac{|d|^2 U}{|c|^2 + |d|^2} \end{pmatrix}, \quad {}^{\text{H}}F^{\downarrow} = \begin{pmatrix} h_a^{\downarrow} + \frac{|a|^2 U}{|a|^2 + |b|^2} & -t \\ -t & h_b^{\downarrow} + \frac{|b|^2 U}{|a|^2 + |b|^2} \end{pmatrix} \qquad (24)$$

and

$$E_{\text{Hartree}}^{\text{H}} = U \frac{(|a|^2 |c|^2 + |b|^2 |d|^2)}{(|a|^2 + |b|^2)(|c|^2 + |d|^2)}. \qquad (25)$$

Having derived the explicit energy expressions for the HMF and NHMF, the solution to these model then consists in finding the stationary points of the energy expressions where

$$\frac{\partial E^{(\text{N})\text{HMF}}(a,b,c,d)}{\partial x} = 0, \quad \text{for } x = a, b, c, d. \qquad (26)$$

The various stationary points obtained are then compared to the exact solutions of the Hubbard model which we construct next.

## IV. EXACT SOLUTION OF THE TWO-SITE HUBBARD MODEL

First we introduce the basis functions of the two-electron space, after eliminating the parallel-spin states we are left with four basis functions for the two-electron system and they are listed in Tab. I. Setting $E_1 = h_a^{\uparrow} + h_a^{\downarrow} + U$, $E_2 = h_b^{\uparrow} + h_b^{\downarrow} + U$, $E_3 = h_a^{\uparrow} + h_b^{\downarrow}$, $E_4 = h_a^{\downarrow} + h_b^{\uparrow}$,

TABLE I. The basis functions of the two-site Hubbard model.

| Index | Basis Function | Description |
|---|---|---|
| $\|\psi_1\rangle$ | $c_{1\uparrow}^\dagger c_{1\downarrow}^\dagger\|0\rangle$ | Both electrons on site 1 |
| $\|\psi_2\rangle$ | $c_{2\uparrow}^\dagger c_{2\downarrow}^\dagger\|0\rangle$ | Both electrons on site 2 |
| $\|\psi_3\rangle$ | $c_{1\uparrow}^\dagger c_{2\downarrow}^\dagger\|0\rangle$ | Electrons on different sites |
| $\|\psi_4\rangle$ | $c_{1\downarrow}^\dagger c_{2\uparrow}^\dagger\|0\rangle$ | Electrons on different sites |

we obtain the Hamiltonian Matrix,

$$H = \begin{pmatrix} E_1 & 0 & -t & t \\ 0 & E_2 & -t & t \\ -t & -t & E_3 & 0 \\ t & t & 0 & E_4 \end{pmatrix}. \qquad (27)$$

The parameter values that we select are $h_a^\uparrow = \frac{1}{4}, h_b^\uparrow = -\frac{1}{4}$, $h_a^\downarrow = h_b^\downarrow = 0$, and $t = 1$. This choice entails that there is no up- and down-spin symmetry, nor is there a left-right mirror symmetry. We purposefully remove spatial- and spin-symmetry which would otherwise lead to simplifications and the elimination of variational degrees of freedom that we want to explore in the present work. In Fig. 2 the eigenvalues of the Hubbard Hamiltonian Eq. 27 are studied as a function of the interaction parameter $U$. The ground-state energy curve shows a non-linear behavior, reaching an asymptotic value where the up-spin electron is localized on the right and the down-spin electron is localized on the left atom. For the asymptote of the first excited states, the placements of the up- and down-spin electron are reversed with respect to the ground state. Interestingly, the energy of this state is almost independent of $U$ and equal to zero. The two high energy states correspond to electrons that are localized on the same atom (left or right) and the energy thus increases linearly. As $U \to \infty$, all eigenstates reduce to a single configuration of Tab. I.

## V. SOLUTIONS OF THE HERMITIAN MEAN-FIELD MODEL

First we consider the MF solution for Hermitian model, i.e., the various stationary points of $E^{\mathrm{HMF}}(a, b, c, d)$, which are obtained with a symbolic mathematics program (Mathematica) [29]. In Fig. 3, the solid curves represent the stationary-state energies as a function of $U$.

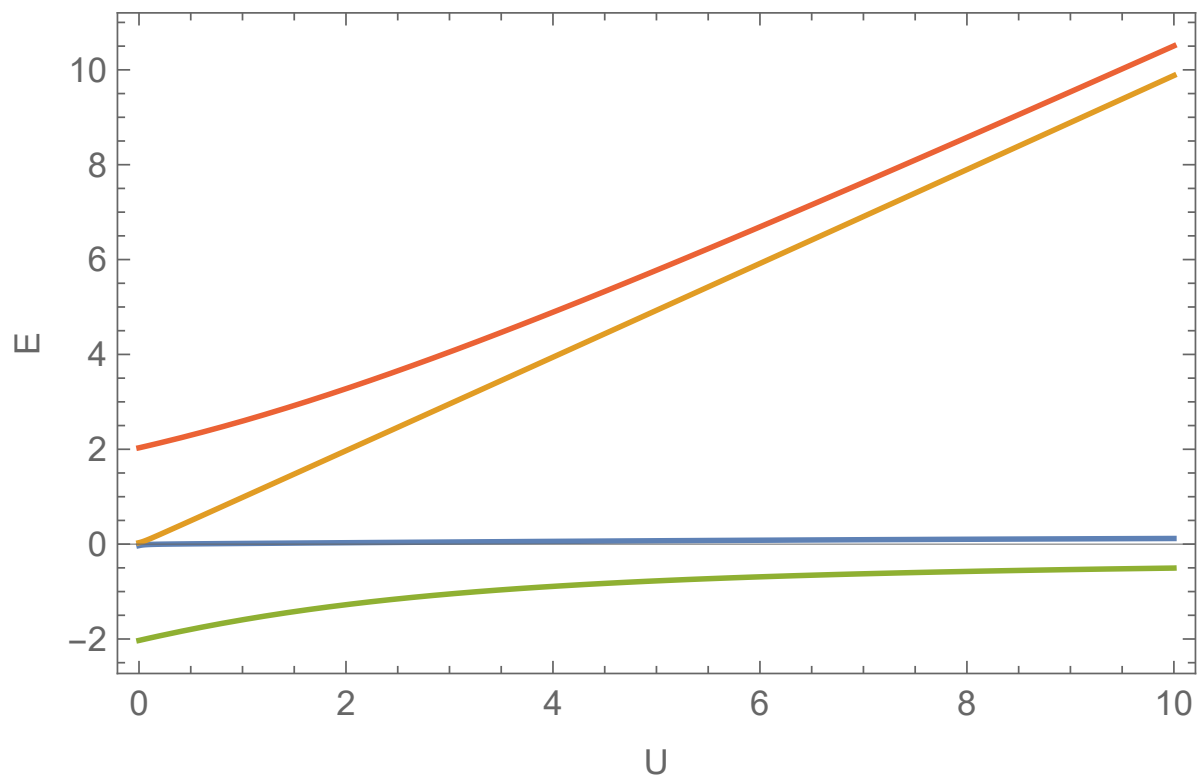

FIG. 2. The exact eigenvalues of the Hubbard Hamiltonian as a function of the on-site repulsion $U$. The two lowest energy states are those describing asymptotic ($U \to \infty$) electron separation, corresponding to $\psi_{3,4}$ in Tab. I. The two high-energy states asymptotically reduce to the ionic configurations $\psi_{1,2}$.

The MF ground-state closely follows the exact solution over the entire range of $U$ values. Asymptotically, the first excited state is also well described within MF, however, as $U$ drops below about 2.8, four MF solutions disappear and the first excited state is no longer represented. The MF solutions can be grouped into two categories, one which behaves non-linearly as a function of $U$, containing a maximum of four solutions, and one which behaves linearly, also containing up to four solutions. An obvious question that arises is what happened to the first excited state in the HMF case as $U$ drops below $\approx 2.8$? By relaxing the condition of Hermiticity, this mystery is resolved in the next section.

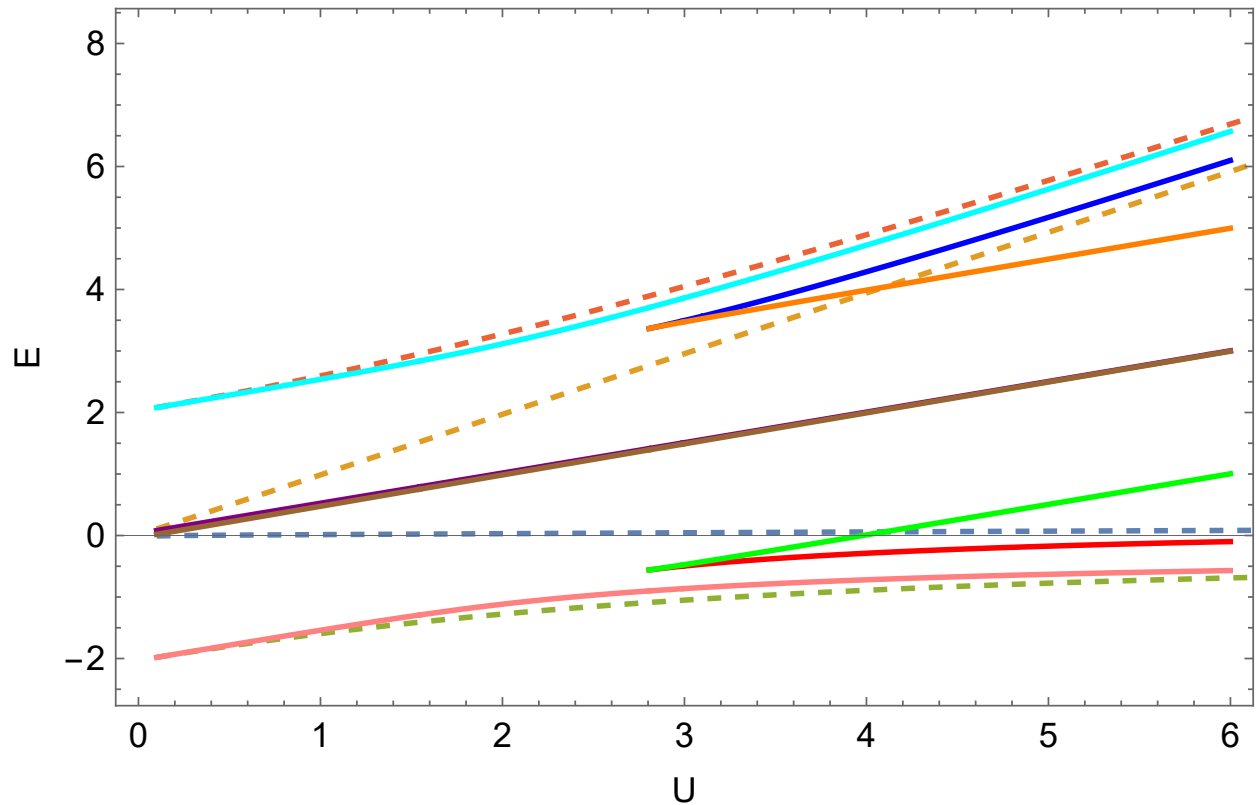

FIG. 3. Plotted are the Hermitian MF solutions of the two-site model (solid lines), they are compared to the exact solutions (dashed lines). There are four MF solutions extending over the entire $U$-range of the plot. Furthermore, there are four solutions that appear for $U \approx 2.8$ and beyond. Two MF solutions, starting at $(U = 0, E \approx 0)$, are almost on-top of each other. Asymptotically, all four exact solutions are recovered by MF ones.

## VI. STATIONARY POINTS OF THE NON-HERMITIAN MEAN-FIELD APPROACH

We consider the NHMF approach and analyze the ensemble of stationary points of $E^{\text{NHMF}}(a, b, c, d)$ which are plotted as a function of $U$ in Fig. 4. Over the entire $U$ range we find eight solutions, where the HMF solutions disappear, the NHMF ones exhibit complex energies, indicating meta-stable states. To analyze the NH solutions, we focus on the first excited state and in particular we ask the questions: What does its continuation into the complex energy plane represent, what is its physical and chemical significance, does it still approximate the first excited state? Here we argue that the first excited state is indeed represented by a complex-energy MF solution. To support this view, we use the orbital co-efficients of NH stationary points to calculate the Hermitian energy, i.e. the physical energy expression $E^{\mathrm{HMF}}(a,b,c,d)$ is evaluated with the stationary points of $E^{\mathrm{NHMF}}(a,b,c,d)$ and the resulting energies are plotted in Fig. 5. The first excited state, represented by the red solid curve, which converges towards the exact first excited state in the asymptotic region, now continues all the way to $U = 0$. It provides a realistic representation of the exact first excited state energy over the entire $U$ range.

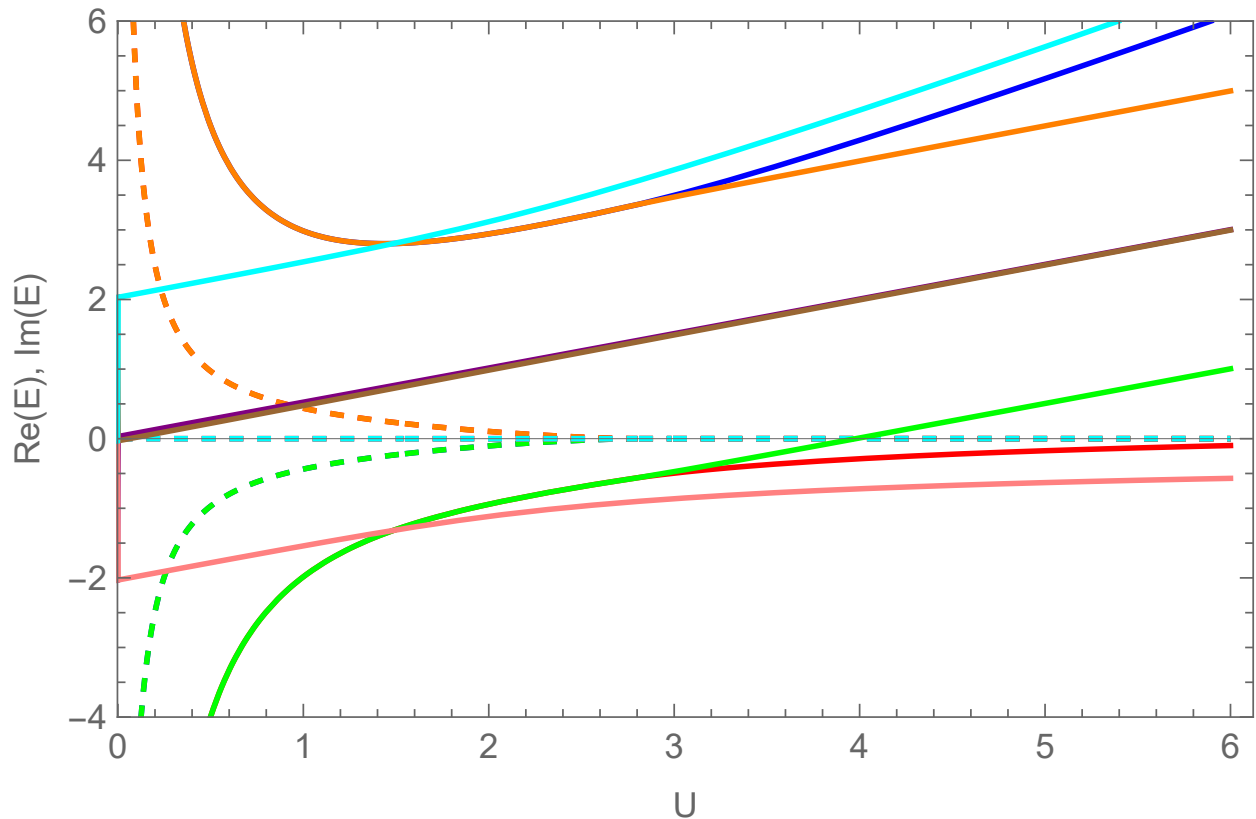

FIG. 4. Real and imaginary parts of the energies obtained with NHMF, plotted as a function of $U$. Real parts are represented by solid curves while the imaginary parts are dotted lines of the corresponding color. There are four states at $U < 2.8$ that have a non-vanishing imaginary part, they are continuations of the four states that disappear in Fig. 3. They can be grouped into two pairs that have identical real energies for $U < 2.8$ and opposing imaginary parts.

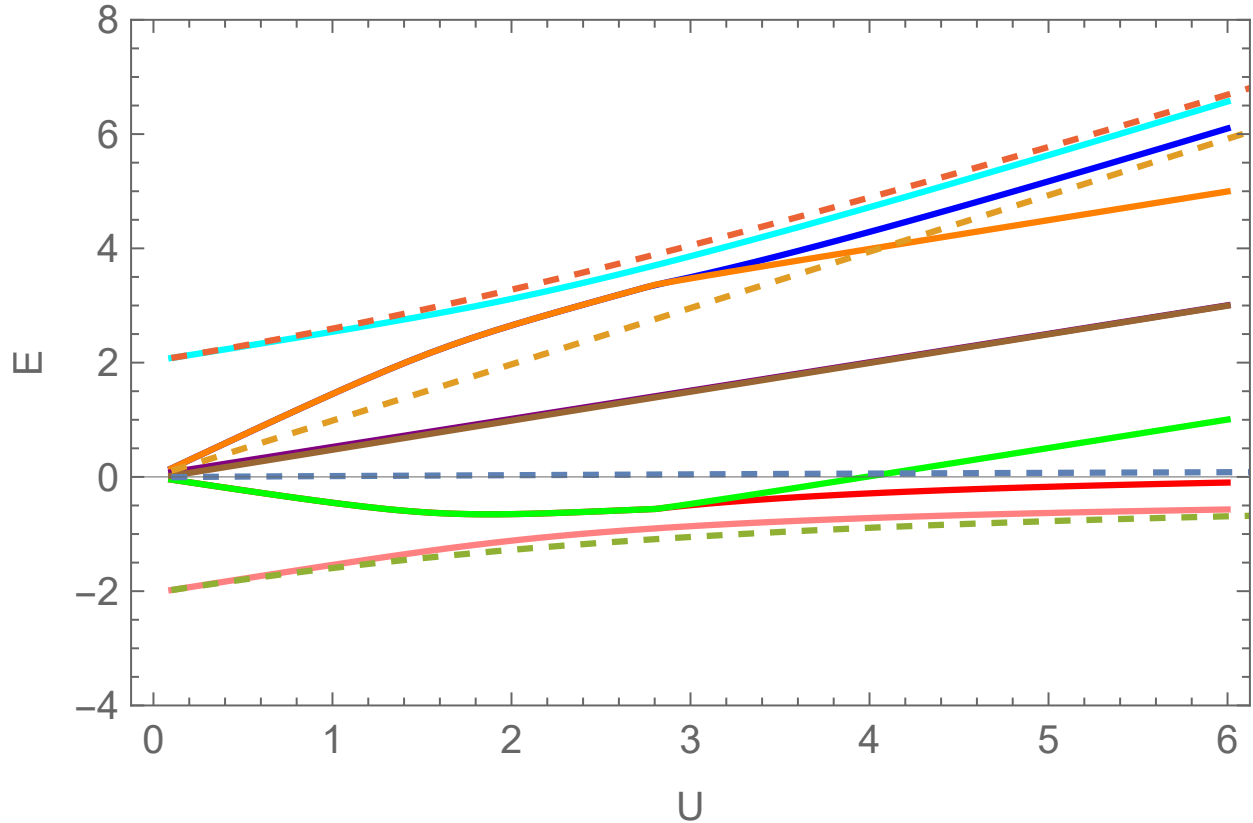

FIG. 5. The energies of the exact solutions of the Hubbard model (dashed curves) are compared to energies obtained the conventional Hermitian MF expression (solid lines) evaluated with the stationary points (orbital coefficients) of NHMF. We argue that the red MF curve (which is covered by the green one for $U < 2.8$) approximates the exact first excited energy (light-blue dashed line). Asymptotically, for $U \to \infty$, these two states (approximate and exact) become identical.

## VII. ANALYSING A NON-HERMITIAN MEAN-FILED SOLUTION

To better understand the significance of the NHMF solutions, in this section, we focus on the first excited state and choosing $U = 1/2$, we investigate it in detail. Its non-Hermitian and Hermitian MF energies are listed in Tab. II, obviously, the NH energy is complex, indicating that the state is meta stable, as noted in Section II. To calculate the physical energy of the NH solution, the Hermitian energy expression is employed and the resulting value is close to the energy of the exact first excited state.

Next we examine the up- and down-spin MF matrices obtained for the first excited state in Tab. III. In an approximate manner, the matrices are $\begin{pmatrix} i & -1 \\ -1 & -i \end{pmatrix}$ and the complex conjugate thereof, respectively. Through numerical studies, where $U$ is reduced to $U = 10^{-4}$, we find that within errors of the order $10^{-4}$

$$\lim_{U \to 0} {}^{\text{NH}}F^{\uparrow} \approx \begin{pmatrix} i & -1 \\ -1 & -i \end{pmatrix}, \quad \text{for the first excited state.} \tag{28}$$

14This is the author's peer reviewed, accepted manuscript. However, the online version of record will be different from this version once it has been copyedited and typeset.
PLEASE CITE THIS ARTICLE AS DOI: 10.1063/5.0272598

TABLE II. Hermitian and non-Hermitian total MF energies, as well as NH occupied-orbital energies. These values are calculated using the orbitals of NH first excited state for $U = \frac{1}{2}$.

| Quantity | Value |
| --- | --- |
| **System Energies** | |
| Non-Hermitian Energy | $-3.993 - 0.970i$ |
| Hermitian Energy | $-0.233 + 0.000i$ |
| Exact Energy | $0.006 + 0.000i$ |
| **Occupied-Orbital Energies** | |
| Orbital 1 | $0.001 + 0.003i$ |
| Orbital 2 | $0.014 + 0.058i$ |

This result is obtained with the Mathematica program [29] and it can be rationalized by noting that for $U \to 0$ the denominator in $\frac{c^2 U}{c^2+d^2}$ and $\frac{d^2 U}{c^2+d^2}$ tends to zero as well, so that the ratio remains finite. This is because for the orbital coefficients $(c, d) \to (1, i)$ and thus $c^2 + d^2 \to 0$ (the orbitals are further discussed below). In the introduction, we describe the SSP method and we provide a very simple example in Fig. 1, it is a puzzling fact that for $U \to 0$, within numerical accuracy, ${}^{\text{NH}}F^{\uparrow}$ reduces to the SSP Hamiltonian in Eq. 2. Similarly, ${}^{\text{NH}}F^{\downarrow}$ reduces to the complex conjugate of the SSP matrix. This analogy facilitates the interpretation of the MF matrices and supports the view that in the NHMF single-particle picture, a given electron is coupled to an environment (consisting of the remaining electrons) and exchanges probability density with this environment. In the limit $U \to 0$ of Eq. 28 the Fock matrices exhibit exceptional points (see, e.g., [30, 31]), where the respective eigenvectors and eigenvalues coalesce. Exceptional points in the context of SSP have been discussed in [32]; here, since only one orbital of each Fock matrix is occupied, no direct consequences for the Hermitian and non-Hermitian energies are observed.

The orbitals obtained by diagonalizing the MF matrices are provided in Tab. IV. Of particular interest are the occupied ones corresponding to the eigenvalues that are close to zero in absolute value. Approximately, the orbitals are $(1, i)$ and $(i, 1)$, and within numerical accuracy these expressions become exact for $U \to 0$. The orbitals describe an electron moving from the left atom to the right and from the right atom back to the left, respectively [4]. Since the velocity of this process is high, implying increased kinetic energy, the corresponding orbital energies are zero. To further elaborate, we note that in continuum quantum mechanics, a forward going electron is described by a plane wave of the form $e^{ikx}$, where the electron velocity is proportional to $k$. In the Hubbard model, $(1, i) = (e^{ikx_1}, e^{ikx_2})$ is a discreet version of such a plane wave where $k = \pi$ and $(x_1, x_2) = (0, 1)$. The conventional binding and anti-binding orbitals, each are a superposition of $(1, i)$ and $(i, 1)$.

TABLE III. Up-spin and down-spin MF (Fock) matrices of the NH first excited state with $U = 1/2$.

| Up-spin mean-field matrix | Down-spin mean-field matrix |
|---|---|
| $\begin{pmatrix} 0.249 + 0.969i & -1.000 \\ -1.000 & 0.251 - 0.969i \end{pmatrix}$ | $\begin{pmatrix} 0.264 - 0.973i & -1.000 \\ -1.000 & 0.236 + 0.973i \end{pmatrix}$ |

TABLE IV. Converged MF matrix eigenvalues and eigenvectors of the NH first excited state with $U = 1/2$.

| Up-spin MF matrix | | Down-spin MF matrix |
|---|---|---|
| **Eigenvalues** | | |
| $0.499 - 0.003i$ | | $0.486 - 0.058i$ |
| $0.001 + 0.003i$ | | $0.014 + 0.058i$ |
| **Eigenvectors** | | |
| $\begin{pmatrix} -0.176 - 0.684i \\ 0.708 + 0.000i \end{pmatrix}$ | | $\begin{pmatrix} 0.728 + 0.000i \\ -0.162 - 0.667i \end{pmatrix}$ |
| $\begin{pmatrix} 0.708 + 0.000i \\ 0.176 + 0.684i \end{pmatrix}$ | | $\begin{pmatrix} 0.162 + 0.667i \\ 0.728 + 0.000i \end{pmatrix}$ |

The interpretation of the orbital energies as quasiparticle energies implies that there are ionization potentials that are nearly zero. This finding is specific to NHMF and not observed in the Hermitian approach where the orbital energies correspond to binding and anti-binding orbitals which are 1.285 and -0.785, respectively for the up-spin orbitals. Very similar values are obtained for the down-spin orbitals. The exact first excited state represents a non-binding one which appears to be best approximated by non-binding complex orbitals (i.e., orbitals of zero energy). The NHMF theory provides a whole new dimension (the imaginary axis) to construct orbitals and to describe physical phenomena. A description of the stationary NHMF state whose Hermitian energy coincides with the one analyzed in this section for $U < 2.8$ (see Fig. 5) is obtained by applying complex conjugation to the results reported in this section.

To conclude the illustration, we describe changes that are observed if the local spin potentials are all equal and zero, i.e., $h_a^\uparrow = h_b^\uparrow = h_a^\downarrow = h_b^\downarrow = 0$ [28]. In this symmetric case, the exact first excited state reduces to the first excited state of triplet symmetry. Upon going from the asymmetric to the symmetric case, in the small $U$ region, the NHMF state remains an excited state of broken spin symmetry that approximates the first excited state of triplet symmetry. In the large $U$ region, there are no NH solutions, neither in the symmetric nor in the asymmetric case. In this region, the first excited state of the asymmetric system becomes degenerate with the ground-state upon elimination of the asymmetry. In the symmetric as in the asymmetric case, for $U \to 0$, the MF matrices of the NH states recover exactly the SSP matrix of Eq. 2 and its complex conjugate. Obviously, in both cases the open-system interpretation of the NH states applies. The total energy of the NH solution is real for the symmetric system, while the individual contributions (kinetic, potential) to the energy are complex, the imaginary parts cancel in the total energy as a consequence of the symmetry.

We would like to emphasize that in the context of KS-DFT, the first- and other excited states are often of interest; starting with the work of Slater about the $\Delta$SCF method [33], formal extensions of ground-state DFT have been developed, e.g., [34, 35] that provide excited-state KS approaches. Similarly, in the time-dependent DFT method for excitation energies, a first excited state that shifts into the complex plane is highly relevant, as it is of course in HF theory. Furthermore, for ground-state DFT, the inclusion of the approximate exchange-correlation energy $E_{\text{XC}}[\rho]$ can lead to solutions that are shifted into the complex plane. For instance, a functional of the form

$$E_{\text{XC}}[\rho] = \int d^3r \left[ A(\mathbf{r})\rho^4(\mathbf{r}) + B(\mathbf{r})\rho^3(\mathbf{r}) + C(\mathbf{r})\rho^2(\mathbf{r}) + D(\mathbf{r})\rho(\mathbf{r}) + E(\mathbf{r}) \right], \quad (29)$$

with appropriately chosen real coefficients $A(\mathbf{r}), B(\mathbf{r}), C(\mathbf{r}), D(\mathbf{r}), E(\mathbf{r})$ can eliminate a sta-tionary point in $E[\rho] - E_{\text{XC}}[\rho]$ which is located at $\rho_0$ and replace it by two stationary points $\rho_1(\mathbf{r})$ and $\rho_1^*(\mathbf{r})$ that are located in the complex-density plane. This is because a term proportional to $\int d^3r (\rho(\mathbf{r}) - \rho_0(\mathbf{r}))^2$ can eliminate the stationary point $\rho_0$ and $\int d^3r (\rho(\mathbf{r}) - \rho_1(\mathbf{r}))^2(\rho(\mathbf{r}) - \rho_1^*(\mathbf{r}))^2$, adds stationary points at $\rho_1(\mathbf{r})$ and $\rho_1^*(\mathbf{r})$; the sum of these two terms an be cast into the form of the right-hand-side of Eq. 29.

## VIII. CALCULATING THE MOLECULAR CONDUCTANCE

Given the emergence of additional NHMF solutions already for the very simple example considered in the previous section, is likely that there are numerous such solutions for more complex systems. One aspect of these NH stationary states is that they add additional effective one-electron levels compared to the conventional MF approaches. In the present context, having already introduced the SSP approach, we use the molecular conductance as a property where the new type of NH solutions of the isolated molecule leads to significant changes in the predictions for the molecular conductance.

The conductance strongly reflects the effective single-particle spectrum, and as already mentioned, the NHMF approach offers a larger such spectrum than the HMF approach, leading to different signatures in the molecular conductance. As described in the introduction, in SSP complex-valued, single-particle potentials are added to the Hamiltonian to account for the external contacts the molecule is connected to. Here we choose a weak coupling of the molecule to the contacts ($\beta = 0.1t$), ensuring that the conduction channels can be related to the NHMF solutions of the isolated system and we set $h_a^\uparrow = \frac{1}{4} + i\beta\frac{1+r}{1-r}, h_b^\uparrow = -\frac{1}{4} - i\beta$, while leaving the remaining parameters unchanged. This choice corresponds to the addition of external contacts and to the creation of a channel for up-spin electron transport. In SSP, the crucial step is the determination of the reflection coefficient $r(E)$ related to the transmission probability through $T = 1 - |r(E)|^2$. For the full Hamiltonian of the Hubbard model, $r(E)$ is obtained with the condition that for a given energy $E$, $\det(H(r(E)) - E) = 0$. The resulting $T(E)$ curve is shifted by the occupied orbital energy to account of the fact that in the MF case (described next), the energy is that of an orbital and not the total energy of the system. For the MF approximation, the parameter $r(E)$ appears in the up-spin MF matrix and in this case it is determined through the condition $\det(^{\text{NH}}F^\uparrow(r(E)) - E) = 0$. This equation is satisfied in addition to the stationarity condition Eq. 26. The NHMF solutions for $r(E)$, obtained with the ground and first excited state, are used to calculate $T(E)$ and compared to the corresponding exact results in Fig. 6. The maxima in $T(E)$ along the perpendicular axis indicate effective single-electron level and the additional stationary points provided by the NHMF approach yield conduction channels that resemble the ones obtained with the exact first excited state solution. This state corresponds to a non-bound diatomic yielding a narrow conduction channel at electron energy equal to zero. In the NHMF case, there are zero-energy orbitals for the first excited state that provide a conduction channel at this energy, albeit of larger width. It is important to realize that these additional conductance channels are an intrinsic feature of the NHMF approach which is the appropriate theory for open-system calculations. This is an observation that has not been appreciated in open-system MF calculations which are usually performed [36] with Hermitian density matrices and real densities.

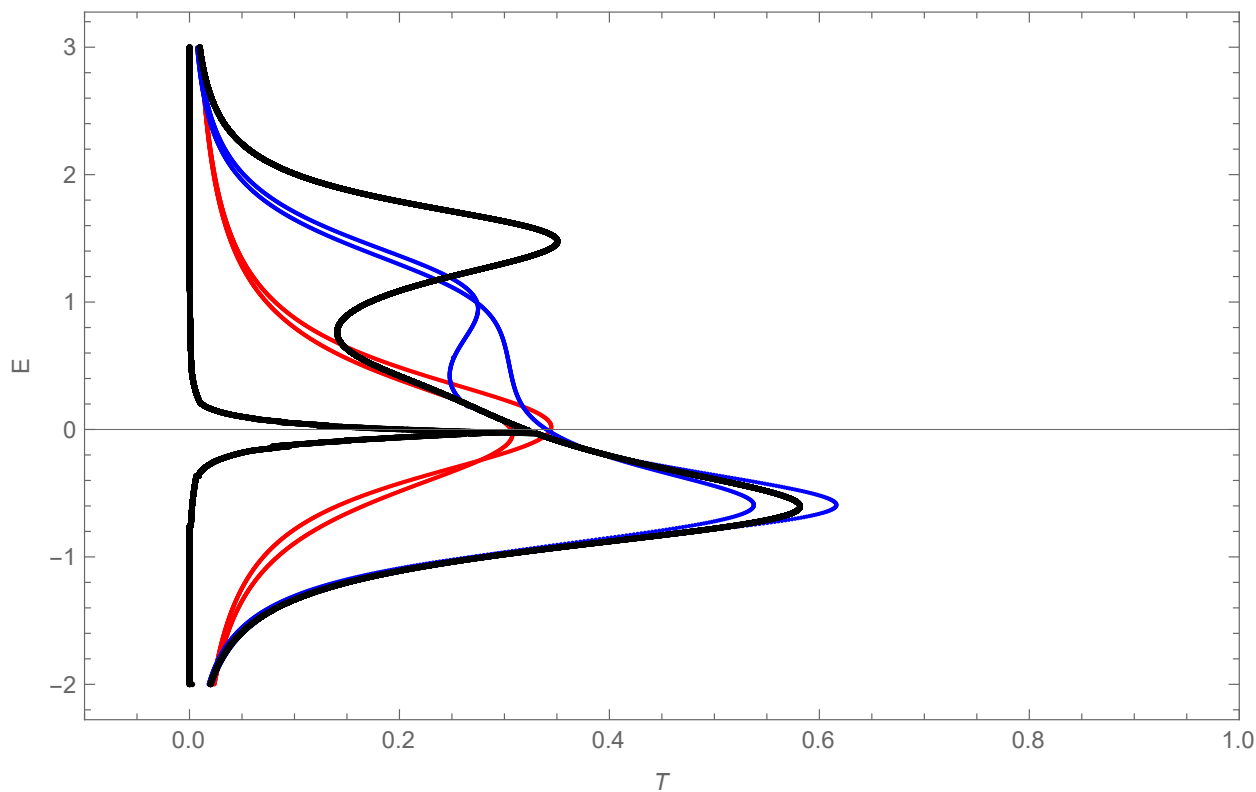

FIG. 6. Transmission probability ($T$) obtained with the exact (black curves) and the MF solutions of the model system. One of the exact curves, the smooth one with two maxima, derives from the ground state. The maxima approximately represent conductance through the binding and anti-binding orbital. The second curve yields a sharp peak around $E = 0$, it derives from the exact first excited state of the isolated system. For the NHMF method, two pairs of curves are displayed, one pair in red and one pair in blue. Pair blue corresponds to the standard independent-particle solution, showing a maximum or shoulder at the energy of the binding and anti-binding orbital. These orbital energies are -1 and +1, respectively. This pair of curves represents a good approximation to the smooth black curve with two maxima. The red pair of NHMF solutions resembles a Gaussian centered at zero. These curves emerge from the NHMF solutions of the isolated system that have complex energies for small $U$ values and that approximate the first excited state. The corresponding orbitals yield additional paths through the molecule that are available at an electron energy around zero. It is important to note that the conventional MF theory does not provide orbitals of zero energy and thus the corresponding pair of curves is not represented at all within HMF. In the MF case, pairs of solutions appear because there is a coupling of the movement in one orbital to the other that results from the imaginary part of $U\rho_i^\uparrow \rho_i^\downarrow$, and this coupling results in pairs of curves.

## IX. CONCLUSION

Self-consistent MF methods are the central approximations of electronic structure theory, with HF and KS being the primary examples. Here we explore a branch of HF and KS where even for isolated systems, within the NH approach, self-consistent solutions are found where the density is complex and the total as well as orbital energies are complex in general, indicating solutions of finite lifetime. There is no a-priory reason to disregard these solutions as approximations to eigenvalues and eigenvectors of the exact Hamiltonian. The Hermitian energy of a stationary point of NHMF might be a better approximation than the energy obtained from the conventional theory. Indeed, we show that already for a very simple example, the conventional MF method does not provide solutions for relevant states over the entire range of $U$ values, whereas the NHMF approach does. If the employed MF approach would yield the exact stationary points for the problem considered, the total energies for an isolated system would of course be real, but in general this isn't the case and the approximate results can be off in the real and in the imaginary direction.

The extension of MF methods to NH density matrices is somewhat similar to the well known transition from restricted to unrestricted calculations, the total energy can be better approximated by less constraints on the orbitals. The NHMF method provides additional solutions for the orbitals and their energies, thus additional sets of quasi particles that can be related to features of the exact solutions. Indeed, calculating the molecular conductance shows that the exact solutions exhibit conduction channels that are only approximated by the NHMF theory, whereas the conventional MF theory fails to capture them. We would like to stress that within the treatment of open-systems, NHMF is the appropriate generalization of MF approaches and the new class of solutions presents an intrinsic feature of that theory.

The present work calls attention to an uncharted space of solutions of the SCF procedure, both in HF and KS. While with the Hubbard model we only consider the simplest non-linear term in the energy functional, namely one involving the density squared, in general there are non-linear terms of any order in an approximation to the exchange-correlation energy of DFT, leading to a vast space of additional solutions that have not been explored. Furthermore, in the examples considered here, the left and right eigenvectors (orbitals) of the MF operator are related by complex conjugation, this is because the MF potential $U\rho^\uparrow\rho^\downarrow$ is diagonal and the MF matrices are complex symmetric (see Eq. 4 and its discussion). In the general case, the MF potential is non-local and the left and right orbitals are different, leading to even more variational degrees of freedom and to more stationary points.

The NH perspective also provides new challenges for the construction of approximations to the exchange-correlation energy and for density functionals in general. In addition to satisfying constraints for real densities, approximations should also do so if continued into the complex-density realm, where the performance of functionals is further challenged. Improvements in functionals of the complex realm should then translate into ameliorations for real densities as well.

## ACKNOWLEDGMENTS

ME would like to acknowledge the hospitality extended to him by Carlo Adamo and Ilaria Ciofini during a stay at l'École nationale supérieure de Chimie de Paris, where some of this work was done. The financial support through The Natural Sciences and Engineering Research Council of Canada (NSERC) grant RGPIN-2022-04971 is gratefully acknowledged.

## AUTHOR DECLARATIONS

**Conflict of interest**

The authors have no conflicts to disclose.

## DATA AVAILABILITY STATEMENT

The data that supports the findings of this study are available within the article.

---

[1] N. Moiseyev, *Non-Hermitian Quantum Mechanics* (Cambridge University Press, 2011).

[2] I. Rotter, A non-Hermitian Hamilton operator in open quantum physics: review, Journal of Physics A: Mathematical and Theoretical **42**, 153001 (2009).

[3] F. Goyer, M. Ernzerhof, and M. Zhuang, Source and sink potentials for the description of open systems with a stationary current passing through, J. Chem. Phys. **126**, 144104 (2007).

[4] M. Ernzerhof, A Simple Model of Molecular Electronic Devices and Its Analytical Solution, J. Chem. Phys. **127**, 204709 (2007).

[5] B. T. Pickup and P. W. Fowler, An analytical model for steady-state currents in conjugated systems, Chemical Physics Letters **459**, 198 (2008).

[6] D. Mayou, Y. Zhou, and M. Ernzerhof, The Zero-Voltage Conductance of Nanographenes: Simple Rules and Quantitative Estimates, Journal of Physical Chemistry C **117**, 7870 (2013).

[7] Y. Zhou and M. Ernzerhof, Open-system Kohn-Sham density functional theory, Journal of Chemical Physics **136**, 094105 (2012).

[8] B. T. Pickup, P. W. Fowler, M. Borg, and I. Sciriha, A new approach to the method of source-sink potentials for molecular conduction, Journal of Chemical Physics **143**, 194105 (2015).

[9] S. Fias and T. Stuyver, Extension of the source-sink potential approach to Hartree-Fock and density functional theory: A new tool to visualize the ballistic current through molecules, Journal of Chemical Physics **147**, 184102 (2017).

[10] B. T. Pickup and P. W. Fowler, A correlated source-sink-potential model consistent with the meir–wingreen formula, The Journal of Physical Chemistry A **124**, 6928 (2020).

[11] P. W. Fowler and B. T. Pickup, A simple model of ballistic conduction in multi-lead molecular devices, Applied Sciences **11**, 11696 (2021).

[12] U. V. Riss and H. D. Meyer, Calculation of resonance energies and widths using the complex absorbing potential method, Journal of Physics B: Atomic, Molecular and Optical Physics **26**, 4503 (1993).

[13] M. Ernzerhof, Density functional theory of complex transition densities, The Journal of Chemical Physics **125**, 124104 (2006), _eprint: https://doi.org/10.1063/1.2348880.

[14] A. Wasserman and N. Moiseyev, Hohenberg–kohn theorem for the lowest-energy resonance of unbound systems, Phys. Rev. Lett. **98**, 093003 (2007).

[15] D. L. Whitenack and A. Wasserman, Density functional resonance theory: Complex density functions, convergence, orbital energies, and functionals, Phys. Rev. Lett. **107**, 163002 (2011).

[16] Y. Zhou and M. Ernzerhof, Calculating the Lifetimes of Metastable States with Complex Density Functional Theory, The Journal of Physical Chemistry Letters **3**, 1916 (2012), _eprint: https://doi.org/10.1021/jz3006805.

[17] A. Ghosal, P. Joshi, and V. K. Voora, Taming negative ion resonances using nonlocal exchange-correlation functionals, The Journal of Physical Chemistry Letters **15**, 5994 (2024), pMID: 38814272, https://doi.org/10.1021/acs.jpclett.4c00717.

[18] J. A. Gyamfi and T.-C. Jagau, A new strategy to optimize complex absorbing potentials for the computation of resonance energies and widths, Journal of Chemical Theory and Computation **20**, 1096 (2024), pMID: 38261549, https://doi.org/10.1021/acs.jctc.3c01039.

[19] C. M. Bender, Introduction to pt-symmetric quantum theory, Contemporary Physics **46**, 277 (2005).

[20] T.-C. Jagau, A. I. Krylov, and N. Moiseyev, Extending quantum chemistry of bound states to electronic resonances, Annu. Rev. Phys. Chem. **68**, 525 (2017).

[21] R. Santra and L. S. Cederbaum, Non-Hermitian electronic theory and applications to clusters, Physics Reports **368**, 1 (2002).

[22] L. S. Cederbaum, W. Domcke, and J. Schirmer, Molecular resonances: A complex scaling approach, Adv. Chem. Phys. **65**, 115 (1996).

[23] J. V. Ortiz, Electron propagator theory: an approach to prediction and interpretation in quantum chemistry, J. Chem. Phys. **125**, 211102 (2006).

[24] H. G. A. Burton and A. J. W. Thom, Holomorphic hartree–fock theory: An inherently multireference approach, Journal of Chemical Theory and Computation **12**, 167 (2016).

[25] H. G. A. Burton, M. Gross, and A. J. W. Thom, Holomorphic hartree–fock theory: The nature of two-electron problems, Journal of Chemical Theory and Computation **14**, 607 (2018).

[26] H. G. A. Burton, A. J. W. Thom, and P.-F. Loos, Parity-Time Symmetry in Hartree-Fock Theory, J. Chem. Theory Comput. **15**, 4374 (2019).

[27] R. A. Zarotiadis, H. G. A. Burton, and A. J. W. Thom, Towards a holomorphic density functional theory, Journal of Chemical Theory and Computation **16**, 7400 (2020).

[28] H. Burton, *Holomorphic Hartree-Fock Theory: Moving Beyond the Coulson-Fischer Point*, Ph.D. thesis, University of Cambridge (2020).

[29] I. Wolfram Research, Mathematica, version 14.2, champaign, IL, 2024.

[30] M. Müller and I. Rotter, Exceptional points in open quantum systems, Journal of Physics A: Mathematical and Theoretical **41**, 244018 (2008), publisher: IOP Publishing.

[31] W. D. Heiss, Some Features of Exceptional Points, in *Non-Hermitian Hamiltonians in Quantum Physics*, edited by F. Bagarello, R. Passante, and C. Trapani (Springer International Publishing, Cham, 2016) pp. 281–288.

[32] M. Ernzerhof, A. Giguère, and D. Mayou, Non-hermitian quantum mechanics and exceptional points in molecular electronics, The Journal of Chemical Physics **152**, 10.1063/5.0006365 (2020).

[33] J. C. Slater, *The Self-Consistent Field for Molecules and Solids: Quantum Theory of Molecules and Solids, Volume 4* (McGraw-Hill, New York, 1974).

[34] A. Görling, Density-functional theory beyond the hohenberg-kohn theorem, Phys. Rev. A **59**, 3359 (1999).

[35] P. W. Ayers, M. Levy, and A. Nagy, Time-independent density-functional theory for excited states of coulomb systems, Phys. Rev. A **85**, 042518 (2012).

[36] K. Stokbro, J. Taylor, M. Brandbyge, and H. Guo, Ab-initio based non-equilibrium green's function formalism for calculating electron transport in molecular devices, in *Introducing Molecular Electronics, Springer Lecture notes in Physics 680* (2005) pp. 117–151.